\documentclass[prl,twocolumn,showpacs, floatfix]{revtex4}
\usepackage{amsmath}
\usepackage{graphicx}
\usepackage{times}
\usepackage{dcolumn}
\usepackage{bm}

\def\be{\begin{equation}}
\def\ee{\end{equation}}
\def\bea{\begin{eqnarray}}
\def\eea{\end{eqnarray}}
\def\bse{\begin{subequations}}
\def\ese{\end{subequations}}

\def\be{\begin{eqnarray}}
\def\ee{\end{eqnarray}}

\begin{document}

\title{$p_{x}+ip_{y}$ superfluid from $s$-wave interactions of fermionic
cold atoms}
\author{Chuanwei Zhang$^{1,2}$, Sumanta Tewari$^{1,3}$, Roman M. Lutchyn$%
^{1,4}$, and S. Das Sarma$^{1}$}
\affiliation{$^{1}$Condensed Matter Theory Center, Department of Physics, University of
Maryland, College Park, MD 20742}
\affiliation{$^{2}$Department of Physics and Astronomy, Washington State University,
Pullman, WA 99164}
\affiliation{$^{3}$Department of Physics and Astronomy, Clemson University, Clemson, SC
29634}
\affiliation{$^{4}$Joint Quantum Institute, Department of Physics, University of
Maryland, College Park, MD 20742}

\begin{abstract}
Two-dimensional ($p_{x}+ip_{y}$) superfluids/superconductors offer a
playground for studying intriguing physics such as quantum teleportation,
non-Abelian statistics, and topological quantum computation. Creating such a
superfluid in cold fermionic atom optical traps using $p$-wave Feshbach
resonance is turning out to be challenging. Here we propose a method to
create a $p_{x}+ip_{y}$ superfluid directly from an $s$-wave interaction
making use of a topological Berry phase, which can be artificially
generated. We discuss ways to detect the spontaneous Hall mass current,
which acts as a diagnostic for the chiral $p$-wave superfluid.
\end{abstract}

\maketitle

\emph{Introduction:} In recent years, the physics of the 2D chiral $p$-wave (%
$p_{x}+ip_{y}$) superfluids has attracted much attention \cite{Nayak}
because of its nontrivial statistical properties \cite{Ivanov} and potential
application in topological quantum computation \cite{Tewari,Zhang}. The
chiral superfluid can also act as a testbed for studying the true quantum
phenomena such as quantum teleportation and violation of Bell's inequality
\cite{Zhang,Tewari2}, which are often masked by the many-body effects in a
macroscopic system. There has been considerable evidence that the symmetry
of the superconducting order parameter in strontium ruthenate (Sr$_{2}$RuO$%
_{4}$) is spin-triplet $p_{x}+ip_{y}$ \cite{Xia}, but the observation of
exotic properties such as quantum half-vortices and non-Abelian statistics
is a serious problem in Sr$_{2}$RuO$_{4}$ because of intrinsic spin-orbit
coupling in the $p$-wave order parameter \cite{Sarma}. With the recent
observations of $p$-wave Feshbach resonances in spin-polarized $^{40}$K and $%
^{6}$Li atoms \cite{Regal,Schunck,Esslinger,Jin}, a $p_{x}+ip_{y}$
superfluid of fermionic cold atoms may also be realizable in the future \cite%
{Melo,Gurarie,Cheng}. Schemes for observing anyonic statistics and
implementing topological quantum computation using vortices in these systems
have recently been proposed \cite{Tewari,Zhang}. A potential advantage of
the cold atom $p_{x}+ip_{y}$ superfluid is that, being
spin-polarized, non-Abelian statistics associated with the Majorana mode in
the vortex cores should be readily observable.

In spite of its promise, because of the short lifetimes of the $p$-wave
pairs and molecules in experiments \cite{Jin}, realizing a chiral $p$-wave
superfluid from the $p$-wave Feshbach resonances \cite%
{Regal,Schunck,Esslinger,Jin} seems, at present, challenging. To circumvent
this problem, in this paper we propose to use the much more commonplace,
attractive $s$-wave Feshbach resonances, coupled with an artificially
generated topological Berry phase \cite{Berry}, to create a $p_{x}+ip_{y}$
superfluid of fermionic cold atoms. The topological Berry phase, in
principle, can be generated in a variety of ways. Here we consider the Berry
phase originating from an effective, artificial spin-orbit coupling of
atoms. In ultra-cold atomic gases, the effective spin-orbit coupling can be
implemented by having the atoms move in spatially varying laser fields \cite%
{Niu,Ruseckas,Zhang2,Clark,Zhu}. Since the $s$-wave Feshbach resonances have
already been successfully used to create $s$-wave superfluids, our method
offers a promising new way to create a topological chiral $p$-wave
superfluid directly from the $s$-wave interactions. We also stress that,
using the methods and ideas described in this paper, it should be possible
to realize more complex, chiral $d$-wave ($d_{x^{2}-y^{2}}+id_{xy}$)
superfluid order parameter starting with $p$-wave attractive interactions.

Once a $p_{x}+ip_{y}$ superfluid is realized in experiments, one natural and
important question is how to observe the chirality of the order parameter.
In the superconducting state of Sr$_{2}$RuO$_{4}$, it has been done through
the observation of a non-zero polar Kerr effect \cite{Xia}, which
demonstrates macroscopic time-reversal symmetry breaking of the $%
p_{x}+ip_{y} $ order parameter. However, such methods cannot be used to
detect the neutral superfluid order parameter, because superfluids, in
contrast to superconductors, do not directly couple to an electromagnetic
field. Here we show that it is possible to detect the spontaneous Hall mass
current, a clear diagnostic of the $p_{x}+ip_{y}$ order parameter, by
coupling the neutral atoms to effective `electric fields' generated by
optical potentials of laser fields. The method to create a $p_{x}+ip_{y}$
superfluid directly from an $s$-wave interaction, coupled with the methods
to observe the spontaneous Hall transverse mass current, gives a complete
description of a promising new way to create and analyze a chiral $p$-wave
superfluid in fermionic optical traps.

$p_{x}+ip_{y}$\emph{\ superfluid from }$s$\emph{-wave interactions:} We
consider $N$ fermionic cold atoms confined to a quasi-two dimensional ($xy$
plane) trap. The atomic dynamics along the $z$ axis are frozen out by
optical traps with a high trapping frequency \cite{Raizen} or an optical
lattice with high potential depths \cite{Porto}. The Hamiltonian of the
system is
\begin{eqnarray}
H &=&\sum_{i=1}^{N}\left[ \frac{\mathbf{p}_{i}^{2}}{2m}-\mu +U\left( \mathbf{%
r}_{i}\right) +h_{0}\sigma _{i}^{z}+H_{i}^{so}\right]  \notag \\
&+&\frac{1}{2}\sum_{i\neq j}V_{\uparrow \downarrow }\left( \mathbf{r}_{i}-%
\mathbf{r}_{j}\right)  \label{Hamiltonian}
\end{eqnarray}%
where $U\left( \mathbf{r}_{i}\right) $ is the external magnetic harmonic
trap potential on the $xy$ plane. We assume that $U\left( \mathbf{r}%
_{i}\right) $ is weak and the system can be taken as spatially uniform. $%
V_{\uparrow \downarrow }\left( \mathbf{r}_{i}-\mathbf{r}_{j}\right) =g\delta
\left( \mathbf{r}_{i}-\mathbf{r}_{j}\right) $ is the attractive $s$-wave
interaction ($g<0$ is the interaction strength) between the atoms with
opposite spins. In this BCS regime, molecules are not energetically
preferred, therefore their number is strongly suppressed and their effects are
negligible. $\mu $ is the chemical potential and $h_{0}$ is an effective
Zeeman field for the atoms. $H_{i}^{so}=\gamma \left( \mathbf{p}_{i}\times
\mathbf{\sigma }_{i}\right) \cdot \mathbf{e}_{z}$ is the Rashba type
effective spin-orbit coupling \cite{Niu}, where $\mathbf{\sigma }_{i}=\left(
\sigma _{i}^{x},\sigma _{i}^{y},\sigma _{i}^{z}\right) $ is a vector whose
components are the Pauli matrices, and $\gamma $ is the spin-orbit coupling
strength.

The spin dependent part $h_{0}\sigma
_{i}^{z}+H_{i}^{so}$ in Eq. (\ref{Hamiltonian}), can be engineered in a variety of ways. For instance,
ultra-cold atoms moving in a 2D spin-dependent hexagonal optical lattice
\cite{Niu}, or in a non-Abelian gauge potential created by spatially varying
laser fields \cite{Zhang2,Clark}, will experience an effective spin-orbit
coupling. The spin $\uparrow $ and $\downarrow $ in the effective
Hamiltonian (\ref{Hamiltonian}) denote the effective spins. Their
definitions depend on the way to create the effective spin-orbit coupling
and already include the spatial dependence of the lasers for generating the
effective spin-orbit coupling \cite{Niu,Zhang2,Clark}. Therefore, the
corresponding laser parameters, such as the intensity and the optical
lattice spacing, do not appear in the effective Hamiltonian (\ref%
{Hamiltonian}) explicitly, although they do affect the parameters $\gamma $ and $h_{0}$. We
note here that our proposal to generate the $p$-wave interactions does not
depend on the specific methods to generate the effective spin-orbit
coupling. Because of that, we have not specified the definitions of the
effective spins in this paper. In the experiments, $h_{0}$, $\gamma $, and $%
g $ can be adjusted by varying the laser parameters and the Feshbach
resonance.

Since we consider only the short range attractive interaction, the Fourier
transform of the two-body interaction is approximately a constant, $V\left(
\mathbf{q}\right) =g$. Performing second quantization we obtain,
\begin{eqnarray}
V_{\mathbf{k}_{1}^{\prime }\mathbf{k}_{2}^{\prime };\mathbf{k}_{1}\mathbf{k}%
_{2}} &=&\left\langle \mathbf{k}_{1\uparrow }^{\prime }\mathbf{k}%
_{2\downarrow }^{\prime }\right\vert V_{\uparrow \downarrow }\left( \mathbf{r%
}_{i}-\mathbf{r}_{j}\right) \left\vert \mathbf{k}_{1\uparrow }\mathbf{k}%
_{2\downarrow }\right\rangle   \notag \\
&=&g\int \frac{d^{2}q}{(2\pi )^{2}}\delta _{\mathbf{k}_{1}^{\prime },\mathbf{%
k}_{1}+\mathbf{q}}\delta _{\mathbf{k}_{2}^{\prime },\mathbf{k}_{2}-\mathbf{q}%
}\left\langle u_{n\mathbf{k}_{1\uparrow }^{\prime }}\!\!\!|u_{n\mathbf{k}%
_{1\uparrow }}\right\rangle   \notag \\
&&\times \left\langle u_{n\mathbf{k}_{2\downarrow }^{\prime }}\!|u_{n\mathbf{%
k}_{2\downarrow }}\right\rangle ,
\end{eqnarray}%
where $\left\vert u\right\rangle $'s are the single particle eigenfunctions
of the one-body part of the Hamiltonian given by Eq.~(\ref{Hamiltonian}),
and $n$ is the band index. (In the presence of optical lattice potentials,
and in the absence of spin-orbit coupling, $\left\vert u\right\rangle $ is simply
the periodic part of the Bloch wavefunction). Specializing to the BCS
reduced Hamiltonian and ignoring the residual interactions,
\begin{equation}
V_{\mathbf{k},\mathbf{k}^{\prime }}=g\left\langle u_{n\uparrow }(\mathbf{k}%
^{\prime })\!\!|u_{n\uparrow }\left( \mathbf{k}\right) \right\rangle
\left\langle u_{n\downarrow }(-\mathbf{k}^{\prime }\!\!)|u_{n\downarrow
}\left( -\mathbf{k}\right) \right\rangle .  \label{Reduced-Interaction}
\end{equation}%
Note that, for slowly varying scattering potentials $V_{\uparrow \downarrow
}\left( \mathbf{r}_{i}-\mathbf{r}_{j}\right) $ (i.e., the Fourier
transformation $V\left( \mathbf{q}\right) $ is nonzero only for small $%
q=\left\vert \mathbf{k}^{\prime }-\mathbf{k}\right\vert $), the
multiplicative factor on the r.h.s. of Eq. (\ref{Reduced-Interaction}) can
be related to the sum of the Berry phases, $\Phi _{\uparrow \left(
\downarrow \right) }=\int_{\mathbf{k}}^{\mathbf{k}^{\prime }}\left\langle
u_{\uparrow \left( \downarrow \right) }\left( \mathbf{k}^{\prime \prime
}\right) |\frac{\partial }{\partial \mathbf{k}^{\prime \prime }}|u_{\uparrow
\left( \downarrow \right) }\left( \mathbf{k}^{\prime \prime }\right)
\right\rangle \cdot d\mathbf{k}^{\prime \prime }$, for the up and the down
spins \cite{Niu2}. However, for the scattering potentials which are not
slowly varying, $V\left( \mathbf{q}\right) $ takes nonzero value even for
large $\left\vert \mathbf{k}^{\prime }-\mathbf{k}\right\vert $ (the
situation considered in this paper), and the extra phases renormalizing the
interaction are the Pancharatnam geometric phases \cite{Pancharatnam}. In
the following, we show that with a Rashba type of spin-orbit coupling, one
can create an effective $p_{x}+ip_{y}$ pairing interaction $V_{\mathbf{k},%
\mathbf{k}^{\prime }}$ from the $s$-wave interaction.

In the presence of Rashba spin-orbit coupling, the one-body part of the
Hamiltonian can be written as,
\begin{equation}
H_{o}=\left(
\begin{array}{ll}
\varepsilon _{\mathbf{k}}-\mu +h_{0} & -\gamma \left( k_{y}-ik_{x}\right) \\
-\gamma \left( k_{y}+ik_{x}\right) & \varepsilon _{\mathbf{k}}-\mu -h_{0}%
\end{array}%
\right)  \label{Rashba-Hamiltonian}
\end{equation}%
where $\varepsilon _{\mathbf{k}}$ is the single particle kinetic energy. The
two bands of the Hamiltonian (\ref{Rashba-Hamiltonian}) corresponding to the eigenvalues $E_{\pm }(\mathbf{k})=\varepsilon _{%
\mathbf{k}}-\mu \pm \sqrt{h_{0}^{2}+\gamma ^{2}k^{2}}$ are separated by a
gap, $2h_{0}$, at $\mathbf{k}=0$. We choose $\mu $ at the middle of the gap
such that the Fermi surface, the locus in $k$-space where $E_{-}(\mathbf{k}%
)=\mu $, lies in the lower spin-orbit band only with the Fermi momentum
denoted by $k_{F}$. We limit to the regime $\gamma k_{F}\gg
h_{0}\gg E_{c}\left( g\right) $, which necessitates the diagonalization of
the spin-orbit coupling energy as the first step, and work with the lower
band only, where $E_{c}\left( g\right) $ is the energy scale of the BCS
pairing gap, determined by the $s$-wave scattering strength.

Using the appropriate eigenfunction of Eq. (\ref{Rashba-Hamiltonian}), it is
straightforward to show
\begin{equation}
V_{\mathbf{k},\mathbf{k}^{\prime }}=gf\left( k,k^{\prime }\right) \exp
\left( i\left( \theta _{\mathbf{k}}-\theta _{\mathbf{k}^{\prime }}\right)
\right) ,  \label{Factor}
\end{equation}%
where $f\left( k,k^{\prime }\right) =k^{\prime }k/N_{-}^{2}\left( k^{\prime
}\right) N_{-}^{2}\left( k\right) (\alpha +\sqrt{\alpha ^{2}+k^{2}})(\alpha +%
\sqrt{\alpha ^{2}+k^{\prime 2}})$, $N_{-}(k)$ is the normalization constant
for the lower-band eigenfunction, $\theta _{\mathbf{k}}$ is the polar angle
in momentum space, and $\alpha =h_{0}/\gamma $.

BCS pairing occurs on the Fermi surface. In the physical regime $\gamma
k_{F}\gg h_{0}$ considered in this Letter, we have $k_{F}\gg \alpha $. In
this limit, one can show
\begin{equation}
V_{\mathbf{k},\mathbf{k}^{\prime }}\approx g\exp \left( i\left( \theta _{%
\mathbf{k}}-\theta _{\mathbf{k}^{\prime }}\right) \right) /4.
\label{Interaction}
\end{equation}%
The interaction in the angular momentum channel $m$ is given by $%
u_{m}\left( \mathbf{k},\mathbf{k}^{\prime }\right) =\frac{1}{2\pi }%
\int_{0}^{2\pi }d\beta V_{\mathbf{k},\mathbf{k}^{\prime }}e^{im\beta }$,
where $\beta =\theta _{\mathbf{k}^{\prime }}-\theta _{\mathbf{k}}$ is the
angle from $\mathbf{k}$ to $\mathbf{k}^{\prime }$. We therefore have
\begin{equation}
u_{1}\left( \mathbf{k},\mathbf{k}^{\prime }\right) \approx g/4.
\label{Effint2}
\end{equation}%
We see that the bare $s$-wave interaction at $m=0$ channel is now completely
replaced by the $p$-wave interaction at $m=1$ angular momentum channel in
the physical regime $\gamma k_{F}\gg h_{0}\gg E_{c}\left( g\right) $.
Consequently, the pairing interaction is renormalized to a separable
interaction in the $p$-wave channel leading to a ground state of a 2D chiral
$p$-wave superfluid with $p_{x}+ip_{y}$ symmetry of the order parameter: $%
\Delta (\bm p)=\Delta _{0}\left( p_{x}+ip_{y}\right) /p_{F}$, where $\Delta
_{0}$ is the energy gap in the excitation spectrum of the superfluid.

The physical origin of the renormalization of the interaction may be
understood through the Berry phase effects of a Rashba type of spin-orbit
coupling. In the presence of a Rashba type of spin-orbit coupling, an atom
evolving adiabatically in the momentum space accumulates a geometric (Berry)
phase associated with the adiabatic change of the momentum $\mathbf{k}$, in
analogy to the Aharanov-Bohm phase acquired by an electron moving in the
real space in the presence of a magnetic field. Here the corresponding
magnetic field in the momentum space is the Berry curvature field
\begin{equation}
\Omega _{\mathbf{k}}=\left[ \nabla _{\mathbf{k}}\times \left\langle u_{%
\mathbf{k}}|i\frac{\partial }{\partial \mathbf{k}}|u_{\mathbf{k}%
}\right\rangle \right] \cdot \mathbf{e}_{z}=\frac{1}{2}\frac{\alpha }{\left(
\alpha ^{2}+k^{2}\right) ^{3/2}}.  \label{flux}
\end{equation}%
The effective \textquotedblleft magnetic flux" passing through the Fermi
disc is $\Phi _{B}=\int \frac{d^{2}k}{\left( 2\pi \right) ^{2}}\Omega _{%
\mathbf{k}}\approx \pi $ as $k_{F}\gg \alpha $, which means a geometric
phase $\beta \Phi _{B}/2\pi $ is obtained for the adiabatic moving of atoms
from $\mathbf{k}$ to $\mathbf{k}^{\prime }$ ( $\beta =\theta _{\mathbf{k}%
^{\prime }}-\theta _{\mathbf{k}}$). This geometric phase is the origin of
the additional phase factor $\exp \left( -i\beta \Phi _{B}/\pi \right) $ in
the interaction $V_{\mathbf{k},\mathbf{k}^{\prime }}$ (Eq. \ref{Interaction}%
) around the Fermi surface and leads to the $p$-wave pairing at $m=1$
channel. Remarkably, if originally the bare interaction is in the $%
p_{x}+ip_{y}$ channel, the Berry phase renormalizes the interaction to the $%
d $-wave channel ($m=2$). In this way a 2D $d_{x^{2}-y^{2}}+id_{xy}$
superfluid should be realizable, which is very difficult to create using the
conventional Feshbach resonance approach.

In experiments, one can choose a suitable attractive interaction regime (BCS
side) so that the pairing gap for the $s$-wave superfluid would be $\sim
\hbar \times 200$ Hz. The laser parameters for generating the effective
spin-orbit coupling should be chosen so that the Zeeman field $h_{0}$ is $%
\sim \hbar \times 1$ KHz. For a typical Fermi energy $E_{F}\sim \hbar \times
1$ KHz \cite{Jin}, the spin-orbit coupling constant should be chosen so that
$\gamma k_{F}\sim \hbar \times 10KHz$, which should be achievable within the
current experimental technology \cite{Niu,Zhang2,Clark}. With these
parameters, we can limit our discussion to the lower spin-orbit energy band
and create a $p_{x}+ip_{y}$ superfluid from the $s$-wave attractive
interaction using the methods described earlier.

\emph{Transverse Hall mass current in 2D }$p_{x}+ip_{y}$\emph{\ superfluid: }%
Neutral atoms in superfluid can interact with laser fields through dipole
interactions \cite{book}. The dipole interaction can provide an optical
potential, whose gradient can be taken as an ``effective electric field" for
the atoms. Here, we study the linear response of a chiral $p_{x}+ip_{y}$
fermionic superfluid subject to such external effective electric fields
which act as a perturbation. The following two types of external effective
electric fields will be considered.

First, we consider an effective electric field, $E_{y}$, applied along the $%
y $ direction, and calculate the transverse response of
the superfluid along the $x$ direction. This transverse response, which
gives rise to a spontaneous Hall mass current, is a clear diagnostic of the
broken time reversal invariance and the associated chirality of the $%
p_{x}+ip_{y}$ order parameter. The transverse Hall current changes the sign
as the chirality of the order parameter is reversed. As such, this mass
current can be used to detect the realization of the chiral $p$-wave
superfluid. In experiments, this effective electric field, $\mathbf{E}=-%
\mathbf{\nabla }V\left( \mathbf{r}\right) $, can be realized by applying a
perturbation potential $V\left( \mathbf{r}\right) =V_{0}\exp \left(
-y^{2}/2\chi ^{2}\right) $ created by a laser beam traveling along the $x$
direction, where $\chi $ is the beam waist of the laser. For simplicity, we
set the temperature $T=0$ and neglect finite temperature effects. We also
assume that the external trap potential is very weak, and neglect the
effects of the spatial inhomogeneity.

The antisymmetric component of the spontaneous Hall conductivity, $\sigma
_{xy}=-\sigma _{yx}$, for the chiral superfluid can be obtained from the
anomalous chiral response coefficient, which leads to (in momentum and
frequency domain) \cite{Lutchyn},
\begin{equation}
\sigma _{xy}(\mathbf{q},\omega )\approx q^{2}/\,2hd\left( 2\omega
^{2}/v_{F}^{2}-q^{2}\right) ,  \label{sigma_xy}
\end{equation}%
in the low frequencies $\omega \ll \Delta _{0}$ region, where $d$ is the
thickness of the superfluid along $z$ direction, $h$ is the Plank constant, $%
v_{F}$ is the Fermi velocity. Since we consider a time independent
perturbation potential, Eq. (\ref{sigma_xy}) can be simplified to $\sigma
_{xy}\approx -1/2hd$.

Equation\ (\ref{sigma_xy}) can be used to calculate the transverse mass
current $j_{x}\left( \mathbf{r},t\right) =\int \int \sigma _{xy}(\mathbf{q}%
,\omega )E_{y}\left( \mathbf{q},\omega \right) e^{i\mathbf{q\cdot r}%
}e^{i\omega t}d\mathbf{q}d\omega $ induced by the longitudinal effective
electric field $E_{y}$. For the potential $V_{0}\exp
\left( -y^{2}/2\chi ^{2}\right) $, this can be simplified as
\begin{equation}
j_{x}\left( \mathbf{r}\right) =V_{0}y\exp \left( -y^{2}/2\chi ^{2}\right)
/2hd\chi ^{2}.  \label{current1}
\end{equation}%
We see that at the peak of the potential, $y=0$, the mass current is zero.
The current flows in opposite directions on the two sides of the potential,
and reaches maximum at $y=\pm \chi $.

In the time-of-flight measurements, such a current would lead to a velocity
of BCS pairs, $v_{x}\left( \mathbf{r}\right) =j_{x}\left( \mathbf{r}\right)
/n\left( \mathbf{r}\right)$, where $n\left( \mathbf{r}\right) $ is the
density of cooper pairs. The different velocities of the atoms on the two
sides yield a larger image along the $x$ direction compared to the
unperturbed case. The enhancement of the size of the image is determined by
the maximum transverse velocity, $v_{x}^{\max }=V_{0}e^{-0.5}/2n\chi h$,
which occurs for the atoms at $y=\pm \chi $. Assuming representative values
for the parameters, $\chi =20$ $\mu m$, $V_{0}/h=100Hz$, the 2D cooper pair
density $n=10^{12}m^{-2}$, we find $v_{x}^{\max }=8\mu m/s$.\textbf{\ }For a
time of flight 500 ms, the enlargement of the image is about 4 $\mu m$,
which should be observable in experiments.

In the experiments, one can also detect the velocity distribution $%
v_{x}\left( y\right) $ directly by the two-photon Raman
transition \cite{book}. The Raman lasers are focused on a local
region and transfer atoms in that region to another hyperfine state $%
\left\vert 3\right\rangle $. Then one can detect atoms at the state $\left\vert 3\right\rangle $ using the time of flight image,
leading to a determination of their velocity distribution.

The second type of effective electric field we consider is generated by a
laser beam propagating along the $z$ direction and centered at $\left(
x,y\right) =\left( 0,0\right) $. The optical potential can be written as $%
V=V_{0}\exp \left( -r^{2}/2\chi ^{2}\right) $, where $r=\sqrt{x^{2}+y^{2}}$.
The effective electric field is now along the radial direction, which leads
to a response current
\begin{equation}
j_{t}\left( \mathbf{r}\right) =V_{0}r\exp \left( -r^{2}/2\chi ^{2}\right)
/2hd\chi ^{2}  \label{current2}
\end{equation}%
along the tangential direction. Such a current forms a close loop around the
center that corresponds to a rotation of the superfluid (see Fig. \ref%
{rotation}). The velocity reaches its maximum at $r=\chi $ and then
decreases on both sides. The direction of the rotation is determined by the
chirality of the superfluid. This phenomenon can be observed by measuring
the local velocity of the atoms using the Raman process. Remarkably, by
applying a non-rotating laser beam, one can create a rotation of the
condensate in a 2D $p_{x}+ip_{y}$ superfluid. Note that the total angular
momentum is conserved in this process. In a $p_{x}+ip_{y}$ superfluid, each
Cooper pair carries a unit of internal angular momentum. In our scheme, this
angular momentum comes from the effective spin-orbit coupling for the atomic
motion. The external non-rotating laser potential produces a density
gradient of the Cooper pairs, leading to the redistribution of the angular
momentum spatially. This redistribution of the angular momentum yields the
tangential current peaking at $r=\chi $, and is the physical origin of the
rotation of the condensate in a 2D $p_{x}+ip_{y}$ superfluid.

We emphasize that there is no antisymmetric transverse mass current in an $s$%
-wave or $p_{x}$-wave superfluid. Thus, the above experiments involving the
transverse currents can serve as clear diagnostic tests for the existence of
the chiral superfluid. In the $p_{x}+ip_{y}$ superfluid, the time reversal
symmetry is broken, which leads to a non-zero Berry phase in the momentum
space. The non-zero Berry phase, absent in the $s$ or $p_{x}$-wave
superfluids, is the physical origin of the nonzero antisymmetric transverse
mass current.
\begin{figure}[t]
\begin{center}
\includegraphics[scale=0.38]{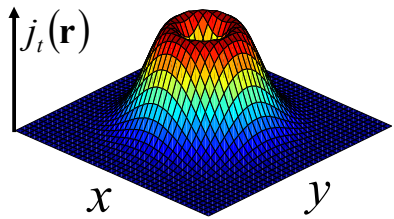}
\end{center}
\caption{Generating a rotating 2D $p_{x}+ip_{y}$ superfluid through a
non-rotating laser propagating along the $z$ direction. The figure
shows the tangential current distribution $j_{t}\left( \mathbf{r}\right) $
for a static laser potential $V\left( \mathbf{r}\right) =V_{0}\exp \left(
-r^{2}/2\protect\chi ^{2}\right) $. The current has the maximum at $r=%
\protect\chi $ and decreases on both sides, where $\protect\chi $ is the
laser beam waist. }
\label{rotation}
\end{figure}

Finally, we note that our proposed method for observing the mass current is
very general. It does not depend on the specific way to generate a $%
p_{x}+ip_{y}$ superfluid. Therefore, if a $p_{x}+ip_{y}$ superfluid can be
generated using other methods (say, using a $p$-wave Feshbach resonance),
our proposed diagnostic methods still apply.

In summary, we have proposed a concrete method to generate a chiral $%
p_{x}+ip_{y}$ cold atom fermionic superfluid by exploiting the
well-established $s$-wave Feshbach resonance and the topological Berry
phases, thereby circumventing the short lifetime issues of $p$-wave
superfluids associated with $p$-wave Feshbach resonance. We have also
proposed techniques for the direct observation of the chirality of the
neutral $p_{x}+ip_{y}$ atomic superfluids in optical traps.

This work is supported by ARO-DARPA. We thank Prof. Qian Niu for stimulating
discussion.

\end{document}